\documentclass[prd,nofootinbib, superscriptaddress,preprint]{revtex4}
\usepackage[T1]{fontenc}
\usepackage{amsmath,amssymb}
\usepackage{epsfig}
\usepackage{dcolumn}
\usepackage{graphicx}
\usepackage[usenames,dvipsnames]{color}
\usepackage{slashed}
\usepackage[colorlinks,citecolor=blue]{hyperref}
\usepackage{pdfpages}
\usepackage{float}
\usepackage{adjustbox}
\usepackage[autostyle]{csquotes}
\usepackage{makecell}

\begin{document}
\title{\boldmath Sensitivity of octant of $\theta_{23}$, CP violation and mass hierarchy in NO$\nu$A with multinucleon and detector effects}
\author{ Paramita Deka \footnote{E-mail: deka.paramita@gmail.com}, Kalpana Bora \footnote{kalpana@gauhati.ac.in},}

\affiliation{Department Of Physics, Gauhati University, Assam, India}

\begin{abstract}

In this work, we investigate how multinucleon enhancement and RPA (Random Phase Approximation) suppression can affect the measurement of three unknown neutrino oscillation parameters - the CP-violating phase $\delta_{CP}$, the octant of the atmospheric mixing angle $\theta_{23}$, and the determination of the mass hierarchy, in the appearance channel of the NO$\nu$A experiment. We include the presence of the detector effect as well in the analysis, which is crucial for capturing realistic experimental scenarios. We also conducted a comparison between the nuclear model Effective Spectral Function (calculated within the RFG model) with and without Transverse Enhancement in terms of sensitivity analysis. It is found that the analysis using our comprehensive model QE(+RPA)+2p2h along with Effective Spectral Function+Transverse Enhancement exhibits significantly enhanced sensitivity compared to the pure QE interaction process, in all the cases.Also, the higher octant of $\theta_{23}$, the lower half plane of $\delta_{CP}$, and the normal mass hierarchy (HO-LHP-NH) exhibit improved sensitivity, enabling a more precise determination of the corresponding parameters. 
Furthermore, it is also noted that improving the performance of the detector also improves the results. Thus, including multinucleon effects and improving detector efficiency have the potential to enhance the capabilities of the NO$\nu$A (and other long baseline) experiment in conducting precise parameter studies.

\end{abstract}
\maketitle

\section{\label{sec:level1}Introduction}

The currently unknown parameters in neutrino oscillations are - the mass hierarchy, the octant of the mixing angle $\theta_{23}$, and the CP violating phase $\delta_{CP}$, and it is widely acknowledged that the presence of degeneracies related to the mass hierarchy-$\delta_{CP}$-octant of $\theta_{23}$ can influence the determination of these parameters. Detecting CP violation in the neutrino sector is a challenging objective for neutrino experiments. Strong evidence of CP violation could be demonstrated by observing an asymmetry in the oscillation rates of muon neutrinos and antineutrinos that undergo a transition into electron neutrinos as in the absence of CP violation, the rates for both are expected to be equal. To conduct this test, both muonic neutrino and antineutrino beams are required. In the context of neutrino interactions, CP violation can occur in the weak force that governs the interaction of neutrinos with matter, and in neutrino oscillation experiments, the phenomenon of CP violation can be observed through differences in the oscillation probabilities between neutrinos and antineutrinos. The discovery of CP violation in neutrino oscillations could help explain the observed matter-antimatter asymmetry in the universe. Modern neutrino experiments do not use free nucleons as the primary target, instead, they use heavy nuclear targets (like carbon, oxygen, argon, etc.), in which complexities regarding nuclear effects are unavoidable. Nuclear effects play an important role and contribute significantly to the incorrect estimation of neutrino energy as they lead to a significant amount of missing energy during Final State Interactions (FSIs). Among all the uncertainties, nuclear effects are considered one of the largest sources of systematic uncertainties in the oscillation analysis of long-baseline (LBL) experiments. Nuclear effects can be broadly divided into two categories - initial-state and final-state effects. Initial-state effects affect the nucleon before the neutrino interactions while hadrons produced by final-state effects influence the outgoing final-state particles before their exit from the nucleus. A true charged current (CC) quasielastic (QE) process is represented as $\nu_{\mu} n\rightarrow\mu^{-}p$. These true interaction processes are accompanied by some other processes where the outgoing proton re-interacts inside the nucleus thus producing $\Delta$ resonance. This $\Delta$ then decays to produce a pion which is then absorbed in the nucleus through FSIs. This can be represented as: $\nu_{\mu} p\rightarrow\mu^{-}p\pi^{+}$ or $\nu_{\mu} p\rightarrow\mu^{-}\Delta^{++}$. Thus the absence of the pion in the final state (called "stuck pion") leads to missing energy and appears as a QE-like event. Therefore because of FSI, a non-QE even may be wrongly identified as QE \cite{Deka:2022idq}. A second complication arises due to the presence of multinucleon events in which the incoming neutrino interacts with, e.g, two nucleons (so-called 2p2h) \cite{Deka:2021qnw, Deka:2022lvy, Martini:2011wp, Martini:2012uc, Martini:2012fa, Benhar:2013bwa, Lalakulich:2012ac}. The neutrino energy reconstructed in such events differs significantly from their true energy value. Though pion production contributes to the background in any QE process, later it has been shown that 2p2h excitations and some other processes also shift the reconstructed energy towards lower energy bins. In MiniBooNE and K2K experiments, it was found that QE contains about 30\% contribution from 2p2h events \cite{Martini:2009uj, Martini:2011wp, Martini:2010ex}. The presence of multinucleon effects modifies the oscillation probabilities and measurements of oscillation parameters and hence can affect the measurements of CP-violating parameters too. Therefore, it is important to understand and properly account for the multinucleon effects in the analysis of neutrino-nucleus interactions. This requires the development of accurate theoretical models of nuclear structure and dynamics, as well as high-precision experimental measurements of neutrino-nucleus cross-sections.

The standard three-flavor neutrino oscillation probability, as expressed in Eq. \ref{prob_app}, depends on six parameters: $\theta_{12}$, $\theta_{23}$, $\theta_{13}$, $\Delta m^{2}_{31}$, $\Delta m^{2}_{21}$, and $\delta_{CP}$. Experimental data from various sources, including solar, atmospheric, reactor, and accelerator experiments, have provided information about each of these oscillation parameters, except for $\delta_{CP}$ \cite{Gonzalez-Garcia:2014bfa, Capozzi:2013csa, Forero:2014bxa}. In LBL neutrino experiments, the $\nu_{\mu}\rightarrow\nu_{e}$ appearance channel is highly sensitive to exploring the CP-violation effect, which remains one of the most challenging problems in neutrino physics today. The oscillation probability for $\nu_{\mu}\rightarrow\nu_{e}$ in the standard three-flavor scenario and constant density approximation can be described by the following expression \cite{DUNE:2020jqi}:\\

$P(\nu_{\mu}\rightarrow\nu_{e})=\sin^{2}\theta_{23}\sin^{2}2\theta_{13}\frac{\sin^{2}(\Delta_{31}-aL)}{(\Delta_{31}-aL)^{2}}\Delta^{2}_{31}+\sin2\theta_{23}\sin2\theta_{13}\sin2\theta_{12}\frac{\sin(\Delta_{31}-aL)}{(\Delta_{31}-aL)}\Delta_{31}$
\begin{equation}
\times \frac{\sin(aL)}{aL}\Delta_{21}\cos(\Delta_{31}+\delta_{CP})+\cos^{2}\theta_{23}\sin^{2}2\theta_{12}\frac{\sin^{2}(aL)}{(aL)^{2}}\Delta^{2}_{21}
\label{prob_app}
\end{equation}

where
\begin{equation}
a=\frac{G_{F}N_{e}}{\sqrt{2}}\approx\pm\frac{1}{3500 km}\frac{\rho}{3.0 g/cm^{3}}
\end{equation}
Here $G_{F}$ represents the Fermi constant and $N_{e}$ is the number density of electrons in Earth's crust. The value of $\rho$ is set to be 2.848 $g/cm^{3}$
\cite{Devi:2020lem}. Additionally, $\Delta m^{2}_{ij}(\equiv m^{2}_{i}-m^{2}_{j})$ represents the difference in mass squares between neutrinos of the $i$-th and $j$-th families, $\Delta_{ij}=(1.267 \Delta m^{2}_{ij} L)/E_{\nu}$, where $L$ is the baseline in kilometers and $E_{\nu}$ is the neutrino energy in GeV. 

For $\nu_{\mu}\rightarrow\nu_{e}$ and $\bar\nu_{\mu}\rightarrow\bar\nu_{e}$ oscillations, the terms $\delta_{CP}$ and $a$ have positive and negative signs, respectively. This leads to a neutrino-antineutrino asymmetry due to both CP violation and the matter effect $a$. The rates of neutrino oscillations, specifically $\nu_{\mu}\rightarrow\nu_{e}$ and $\bar\nu_{\mu}\rightarrow\bar\nu_{e}$, are influenced by the mass hierarchy as neutrinos travel through the Earth in contrast to when they propagate through a vacuum. The presence of coherent forward scattering with electrons in the Earth's crust amplifies the $\nu_{\mu}\rightarrow\nu_{e}$ transition while reducing the occurrence of $\bar\nu_{\mu}\rightarrow\bar\nu_{e}$ in the normal hierarchy (NH). Conversely, this amplification and suppression are reversed in the inverted hierarchy (IH). This phenomenon, known as the matter effect \cite{Wolfenstein:1977ue}, modifies the oscillation probabilities for the NO$\nu$A experiment by approximately 20\% \cite{NOvA:2021nfi}. The coupling between the mass state $\nu_{3}$ and the neutrino states $\nu_{\mu}$ and $\nu_{\tau}$ is primarily determined by the angle $\theta_{23}$. When $\theta_{23}=\pi/4$, called maximal mixing,\cite{Harrison:2002et}, $\nu_{\mu}$ and $\nu_{\tau}$ are equally linked to $\nu_{3}$. In the case of non-maximal mixing, $\theta_{23}$ can exist in either the higher octant (HO) with $\theta_{23}>\pi/4$, or the lower octant (LO) with $\theta_{23}<\pi/4$. 

A non-zero value of $\delta_{CP}$, other than $0^{\circ}$ and $\pm180^{\circ}$, would indicate CP violation in the lepton sector, and $\delta_{CP}=\pm 90^{\circ}$ corresponds to maximum CP violation. It is often convenient to divide the parameter space into the lower half-plane (LHP) with $-180^{\circ}<\delta_{CP}<0^{\circ}$ and the upper half-plane (UHP) with $0^{\circ}<\delta_{CP}<180^{\circ}$. The appearance channel $P_{{\mu}\rightarrow e}$ often known as the golden channel can measure all three unknown parameters. In LBL experiments, the measurement of neutrino oscillation parameters is a challenging task due to the presence of degeneracies. These degeneracies arise because different combinations of parameter values can lead to the same oscillation probability \cite{Agarwalla:2021bzs, Barger:2001yr, Minakata:2004pg}. As a result, determining the true values of the parameters becomes complicated and lacks unambiguous resolution. In Ref. \cite{Capozzi:2021fjo}, the authors observe a preference at a significance level of 1.6$\sigma$ for $\theta_{23}$ in the LO compared to the secondary best-fit in the HO. Assuming NH, they derive a best-fit value of $\sin^{2}\theta_{23}$ = 0.455 in the LO, and at a significance of approximately 1.8$\sigma$, they disfavor maximal $\theta_{23}$ mixing. In Ref. \cite{deSalas:2020pgw}, de Salas et al. find the best-fit value in the HO to be around $\sin^{2}\theta_{23}\sim$ 0.57, assuming NH. On the other hand, Capozzi et al. \cite{Capozzi:2021fjo} and Esteban et al. \cite{Esteban:2020cvm} obtain the best-fit value around $\sin^{2}\theta_{23}\sim$ 0.45 in the LO. These degeneracies make it difficult to uniquely determine the specific values of the parameters. More advanced experimental techniques and additional data are needed to resolve these degeneracies and obtain more accurate measurements of $\delta_{CP}$, the neutrino mass hierarchy, and the octant of $\theta_{23}$ \cite{Bora:2014zwa}.

In this work, we investigate the influence of the multinucleon enhancement and RPA suppression on the determination of the CP-violating phase $\delta_{CP}$, the octant of the atmospheric mixing angle $\theta_{23}$, and the determination of the mass hierarchy in the appearance channel of the NO$\nu$A experiment. In this work, we have used the Effective Spectral function (ESF) model as the nuclear model (calculated within the RFG model), along with the Transverse Enhancement (TE) model \cite{Bodek:2011ps}. We have considered two scenarios: one ESF with TE model and another ESF without TE and studied the difference between these two models in sensitivity analysis of octant of $\theta_{23}$, $\delta_{CP}$ and mass hierarchies. We also include the presence of detector effects of NO$\nu$A experiment, which is an important consideration for realistic experimental scenarios. We consider the multinucleon enhancement and CP violating phase ($\delta_{CP}=\pm 90^{\circ}$) at the same time to know their impact on sensitivity analysis. As seen in our works \cite{Deka:2021qnw, Deka:2022lvy} the RPA suppression and multinucleon enhancement play a significant role in shaping the oscillation probabilities and can have a substantial impact on the determination of the neutrino oscillation parameters, we have also considered the pure QE interaction process in octant and mass hierarchy (MH) sensitivity to observe the deviation from pure QE process to QE+multinucleon enhancement+RPA suppression interaction process. Throughout our study, we have considered the interplay between the multinucleon effect, the detector effect, and the relevant oscillation parameters to gain insights into their combined impact on the precision of our measurements. 

The paper is structured as follows: in section \ref{sec:1}, we briefly discuss the NO$\nu$A experiment, and section \ref{sec:2} provides an overview of the physics principles and simulation details employed in our study. In section \ref{sec:3}, we present the results of our analysis and engage in a comprehensive discussion. Specifically, we focus on the $\nu_{\mu}+\bar\nu_{\mu}$ channel, investigating the effects on various neutrino oscillation parameters, as well as the impact on QE(+RPA)+2p2h and pure QE interaction processes, along with realistic NO$\nu$A detector effects. Finally, in section \ref{sec:4}, we summarize our findings and provide a concluding remark on our work.

\section{The NO$\nu$A Experiment}
\label{sec:1}

The NO$\nu$A (NuMI Off-Axis $\nu_{e}$ Appearance) experiment \cite{NOvA:2016vij} focuses on neutrino oscillations, specifically measuring the probability of $\nu_{\mu}$ disappearance P($\nu_{\mu}(\bar\nu_{\mu})\rightarrow\nu_{\mu}(\bar\nu_{\mu}$)) and $\nu_{e}$ appearance P($\nu_{\mu}(\bar\nu_{\mu})\rightarrow\nu_{e}(\bar\nu_{e})$). To achieve this, two functionally identical detectors are utilized: the 290-ton Near Detector (ND) and the 14-kton Far Detector (FD). Both detectors are positioned off the central beam axis, with the Far Detector placed 14.6 milliradians off-axis. This off-axis configuration enables the narrow neutrino energy flux to peak around 2 GeV, near the first oscillation maximum driven by $\Delta m^{2}_{32}=2.5\times10^{-3} eV^{2}$, thereby enhancing the oscillation probability in the $\nu_{e}$ appearance channel. The NuMI (Neutrinos at the Main Injector) beam \cite{Adamson:2015dkw}, generated at the Fermi National Accelerator Laboratory, serves as the source of neutrinos for the experiment. The ND is positioned at a distance of 1 km from the source and is located 105 m underground. NO$\nu$A measures oscillations by comparing the un-oscillated energy spectra captured by the ND and the oscillated spectra observed by the FD, situated near Ash River, Minnesota, approximately 810 km away from the production target. The NuMI beam is generated by directing 120 GeV protons from the Main Injector at a fixed graphite target, which produces pions and kaons. These particles are then focused into a narrow beam using magnetic horns, and further decay into muons, anti-muons, and their associated neutrinos. Muons and anti-muons are removed from the decay pipe by a 240-meter-thick rock wall.
Both the ND and FD are composed of planes made from extruded polyvinyl chloride (PVC) cells. The flavor composition of the NuMI beam consists of 97.5\% $\nu_{\mu}$, 1.8\% $\bar\nu_{\mu}$, and 0.7\% $\nu_{e}+\bar\nu_{e}$. The primary objectives of the NO$\nu$A experiment are to precisely measure the mixing angle $\theta_{23}$ and determine its octant, determine the neutrino mass hierarchy, and investigate the CP-violating phase $\delta_{CP}$ in the lepton sector. The $\nu_{\mu}(\bar\nu_{\mu})$ disappearance channel is utilized to measure $\Delta m^{2}_{23}$ and $\sin^{2}\theta_{23}$, while the $\nu_{e}(\bar\nu_{e})$ appearance channel allows for measurements of the mass hierarchy, $\theta_{13}$, $\theta_{23}$, and $\delta_{CP}$.

\section{Physics and Simulation Details}
\label{sec:2}

In the analysis of the $\nu_{\mu}$ ($\bar\nu_{\mu})$ appearance channel, the $\nu_{\mu}$($\bar\nu_{\mu})$-CC interaction channel is utilized as the signal in NO$\nu$A detectors. In these detectors, $\nu_{\mu}$-CC interactions are identified by detecting long muon tracks and any associated hadronic activity at the vertex. The un-oscillated spectra from the NuMI beam are initially measured at the ND. These ND spectra are then extrapolated to predict the spectra at the FD. For a detailed description of the extrapolation technique used in this study, refer to our previous work \cite{Deka:2021qnw, Deka:2022lvy}.

To analyze the sensitivity of the neutrino oscillation parameters $\theta_{23}$ and $\Delta m^2_{32}$, we employ the Feldman-Cousins method \cite{Feldman:1997qc} to calculate the confidence level allowed in the parameter space \cite{Deka:2021qnw}. We generate a sample of 1 million CC $\nu_{\mu}$ and $\bar\nu_{\mu}$ events on a carbon target in the energy range of 0-5 GeV using the NO$\nu$A-ND neutrino and antineutrino flux. The simulation is performed using the GENIE (Generates Events for Neutrino Interaction Experiments) v3.0.6 neutrino event generator \cite{Andreopoulos:2009rq}. GENIE is currently employed by several neutrino baseline experiments, including Miner$\nu$a \cite{MINERvA:2019ope}, MINOS \cite{Adamson:2007gu}, MicroBooNE \cite{Chen:2007ae}, T2K \cite{Abe:2018wpn}, and the NO$\nu$A experiment. We consider four interaction processes: QE scattering, resonances from $\Delta$ decay and higher resonances, nucleon-nucleon correlations (2p2h), and deep inelastic scattering (DIS) in both neutrino and antineutrino modes. We focus on simulated events with no pions in the final state. The QE scattering in GENIE is modeled using the Llewellyn-Smith model \cite{LlewellynSmith:1971uhs}. Resonance processes are implemented according to the formalism attributed to the Berger and Sehgal model \cite{Berger:2007rq}. Furthermore, to incorporate the effect of long-range nuclear charge screening due to random-phase approximation correlations, which also modifies the kinematics of QE interactions, we simulate the QE interactions with the Nieves et al. model \cite{Nieves:2004wx}. These effects significantly suppress the QE interaction process at low invariant four-momentum transferred to the nucleus ($Q^{2}$) and slightly enhance it at higher $Q^{2}$, relative to the RFG prediction. The DIS interaction, categorized as a non-resonant process in GENIE, is implemented following the method of Bodek and Yang \cite{Bodek:2002ps}. Two-nucleon knockout (2p2h) events are simulated using the Valencia model developed by Nieves et al. \cite{Nieves:2011pp}. Various approaches exist for modeling the nuclear ground state, and ongoing efforts within GENIE aim to improve these models. Currently, GENIE represents the nuclear ground state using a spectral function that describes the momentum distribution and removal energy of nucleons participating in a lepton-nucleus interaction. In the default historical model employed since GENIE v2, the initial nucleon momentum follows the relativistic Fermi gas (RFG) treatment introduced by Bodek and Ritchie \cite{Bodek:1981wr}. This version of RFG features non-interacting nucleons up to the Fermi momentum, determined through inclusive electron scattering. The ESF model used in GENIE is based on the RFG model, which assumes that the nucleons inside the nucleus behave like a gas of free, non-interacting particles with a Fermi momentum that depends on the nuclear density. However, the RFG model has limitations, as it neglects the effects of nuclear correlations, meson exchange currents, and other complex nuclear dynamics that are important at higher energies and in heavier nuclei. To overcome these limitations, the effective spectral function model incorporates additional phenomenological parameters that are tuned to match experimental data, such as the nuclear binding energy, the nucleon momentum distribution, and the strength of the nucleon-nucleon correlations. The TEM is used to describe the transverse momentum distribution of the struck nucleon inside the nucleus. This model takes into account the effects of final-state interactions between the struck nucleon and other nucleons in the nucleus, which can lead to a broadening of the transverse momentum distribution. The transverse enhancement model is particularly important for high-energy neutrino interactions, where the struck nucleon can have a large momentum and can interact with multiple nucleons in the nucleus. By including the transverse enhancement model, GENIE can accurately predict the transverse momentum distributions of the final state particles in these high-energy interactions. The TEM incorporates a modified transverse form factor to enhance the strength at higher energy loss. When coupled with the ESF as the nuclear model, the TEM effectively introduces Meson Exchange Currents. This combination provides a comprehensive prescription that aligns well with a diverse set of electron scattering data.

The energy of a CC neutrino, which scatters off a nuclear target and produces 'm' mesons while knocking out 'n' nucleons, can be reconstructed using the calorimetric approach as \cite{Deka:2022lvy}:
\begin{equation}
E_{\nu}^{\text{cal}} = E_{\text{lep}} + \sum_{i}(E_{p} - M) + \epsilon_{n} + \sum_{m}E_{m}
\end{equation}
In this equation, $E_{\text{lep}}$ represents the energy of the outgoing lepton, $E_{p}$ and $M$ denotes the energy of the i-th knocked-out nucleon and the mass of the nucleon (target nucleus), respectively. The term $\epsilon_{n}$ corresponds to the single-nucleon separation energy of the outgoing nucleons, which we set to a fixed value of 25 MeV for the carbon target in both neutrino and antineutrino reconstruction methods. Lastly, $E_{m}$ represents the energy of the m-th produced meson.

We have taken into account, the detector response and efficiency, which introduce smearing effects on the measured energies compared to their true values due to finite detector resolution. This leads to a non-zero probability for an event with a true energy $E_{\text{true}}$ to be reconstructed with a different energy $E_{\text{rec}}$. These probabilities are represented by a set of migration matrices. In our study, we consider realistic specifications for the NO$\nu$A experiment, including a selection efficiency of 31.2\% (33.9\%) for $\nu_{\mu}$ ($\bar{\nu}_{\mu}$) events. For demonstration and comparison, we also analyze the case of 80\% detector efficiency. We use a muon energy resolution of 3.5\% and a hadron energy resolution of 25\% \cite{NOvA:2017ohq}, resulting in an overall energy resolution of 7\% for $\nu_{\mu}$-CC events in both detectors.

\section{Results and Discussion}
\label{sec:3}

\begin{figure*}
\centering\includegraphics[width=15cm, height=8cm]{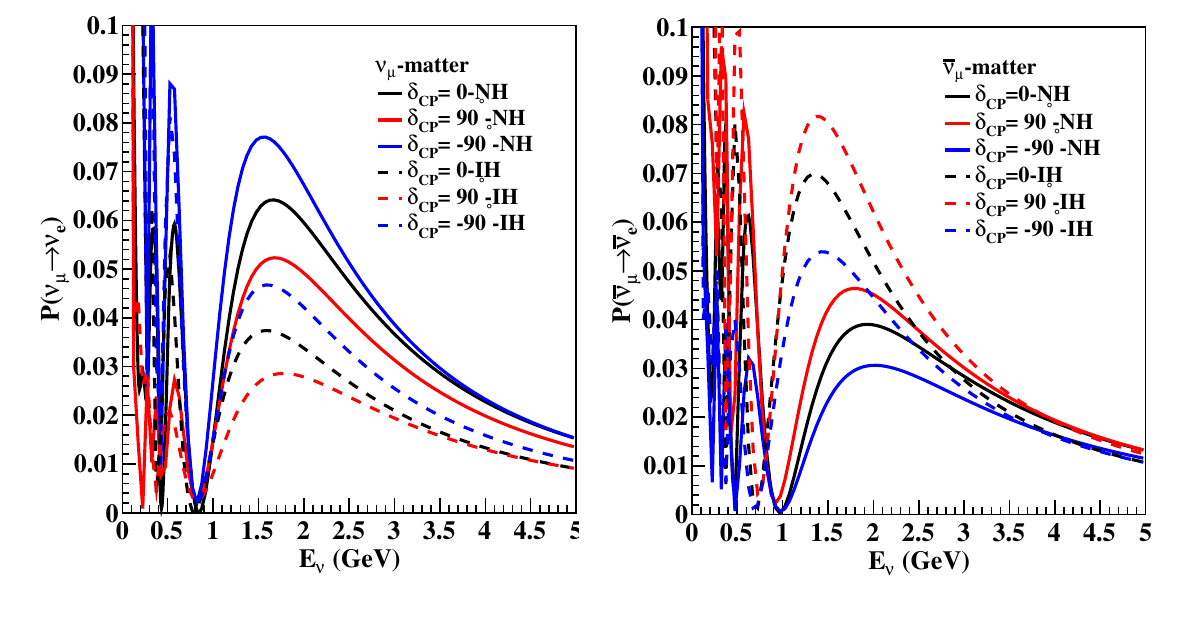}
\caption{$P(\nu_{\mu}\rightarrow\nu_{e})$(solid line) and $P(\bar\nu_{\mu}\rightarrow\bar\nu_{e})$(dashed line) vs neutrino energy for 810 km baseline of NO$\nu$A. Variation of $\delta_{CP}$ leads to variation in oscillation probability.} 
\label{chapter1}
\end{figure*}

Using the details given above, and the technique explained in our papers \cite{Deka:2022idq, Deka:2021qnw, Deka:2022lvy}, we do the sensitivity analysis for CPV phase measurement, octant of $\theta_{23}$ and mass hierarchy, with multinucleon effects, and the results are shown in Figs. (1-8). In the left panel of Fig. \ref{chapter1}, we have presented the variation of $P_{{\mu}\rightarrow e}$ as a function of $E_{\nu}$ for both NH and IH, for $\delta_{CP}$ at $-90^{\circ}$, $0^{\circ}$, and $90^{\circ}$. Generally, the values of $P_{{\mu}\rightarrow e}$ are higher for NH and lower for IH and vice versa for $P_{{\bar\mu}\rightarrow \bar e}$, which directly arises from the matter effect parameter `a'. In the case of neutrino (left panel), for both NH and IH, the curve corresponding to $\delta_{CP}=+90^{\circ}$ appears at the lowest position within the band, while the curve for $\delta_{CP}=-90^{\circ}$ is located at the highest position. This behavior can be easily understood by referring to Eq. \ref{prob_app}. When approaching the oscillation maximum, with $\Delta_{31}\simeq90^{\circ}$, the term $\cos(\Delta_{31}+\delta_{CP})$ becomes +1 for $\delta_{CP}=-90^{\circ}$ and -1 for $\delta_{CP}=+90^{\circ}$. In the right panel of Fig. \ref{chapter1}, we have displayed the corresponding antineutrino probabilities. Here, $P_{{\bar\mu}\rightarrow\bar e}$ is higher for IH and lower for NH due to the sign reversal of `a'. Since $\delta_{CP}$ is reversed for antineutrinos, the upper curves are defined by $\delta_{CP}=+90^{\circ}$, while the lower curves correspond to $\delta_{CP}=-90^{\circ}$. 

Based on Fig. \ref{chapter1}, we can establish the concept of a favorable half-plane for each hierarchy. Assuming NH represents the true hierarchy, when $\delta_{CP}$ is within the LHP (ranging from $-180^{\circ}$ to $0^{\circ}$), all the curves for $P_{{\mu}e}$ (NH, $\delta_{CP}$) must lie above the set of curves for $P_{{\mu} \rightarrow e}$ (IH, $\delta_{CP}$) \cite{Ghosh:2015ena, Prakash:2012az}. In the case of antineutrinos, $P_{{\bar\mu}\rightarrow\bar e}$ (NH, $\delta_{CP}$) will be much lower than $P_{{\bar\mu}\rightarrow\bar e}$ (IH, $\delta_{CP}$). In this scenario, the NH can be determined solely by NO$\nu$A. Therefore, we refer to the LHP as the favorable half-plane for NH. Similar reasoning applies if IH is the true hierarchy and $\delta_{CP}$ lies within the UHP. Hence, UHP is the favorable half-plane for IH. Consequently, we have two possible combinations to choose from: (NH, LHP) or (IH, UHP). The NO$\nu$A experiment can successfully determine the neutrino mass hierarchy in two favorable scenarios \cite{Ghosh:2015ena}:

\begin{itemize}
\item When the hierarchy is NH and $\delta_{CP}$ lies in the LHP ($-180^{\circ}\leqslant\delta_{CP}\leqslant0^{\circ}$).
\item When the hierarchy is IH and $\delta_{CP}$ lies in the UHP ($0^{\circ}\leqslant\delta_{CP}\leqslant180^{\circ}$).
\end{itemize}

\subsection{Octant Sensitivity}

\begin{figure*}
\centering\includegraphics[width=17cm, height=11cm]{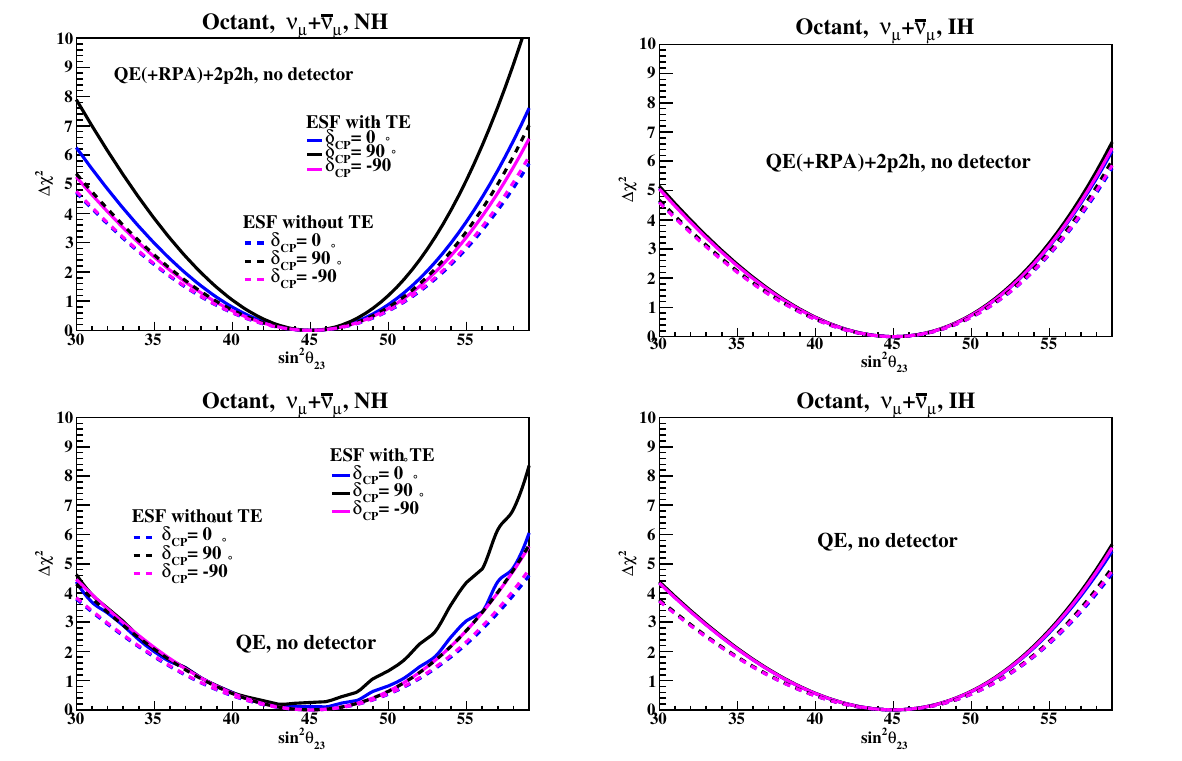}
\caption{The octant sensitivity is compared for various true values of $\delta_{CP}$, considering both QE(+RPA)+2p2h (top panel) and QE (bottom panel) interaction processes, without accounting for the detector effect. The ESF with TE is represented by solid lines, while the ESF without TE is indicated by dashed lines. The analysis is conducted for both the NH in the left panel and the IH in the right panel, focusing on the $\nu_{\mu}$+$\bar\nu_{\mu}$ mode. The figure also highlights the capability of the NO$\nu$A experiment to establish a non-maximal $\theta_{23}$ at $2\sigma$ ($\Delta\chi^{2}=4$) and $3\sigma$ ($\Delta\chi^{2}=9$) confidence levels respectively.}
\label{chapter2}
\end{figure*}

The octant sensitivity of an experiment is determined by comparing the sensitivity of the experiment to measure the correct octant in the true spectrum with the incorrect octant in the test spectrum. In Fig. \ref{chapter2} we show $\Delta\chi^{2}$ as a function of true $\sin^{2}\theta_{23}$ without considering the detector effect for QE(+RPA)+2p2h (top panel) and QE (bottom panel) interaction processes, where for each true value of $\sin^{2}\theta_{23}$, we consider test values of $\sin^{2}\theta_{23}$ in the opposite octant including $\sin^{2}\theta_{23}$(test)=0.5 in the fit. To calculate the octant sensitivity we simulate the data for a representative value of true $\theta_{23}$ belonging to LO (HO) and test it by varying $\theta_{23}$ in the opposite octant i.e., HO (LO). The figure includes three curves: the blue line represents the sensitivity analysis for $\delta_{CP}= 0^{\circ}$, the black line corresponds to $\delta_{CP}= 90^{\circ}$ and the magenta line represents the analysis for $\delta_{CP}= -90^{\circ}$. The solid line represents the ESF+TE and the dashed line ESF without the TE effect. From the results shown in this Fig. \ref{chapter2}, it is observed that:

\begin{itemize}
\item The inclusion of the complete model, which incorporates QE(+RPA)+2p2h (multinucleon enhancement+RPA suppression), demonstrates significantly improved octant sensitivity compared to the pure-QE interaction process. This highlights the crucial role of multinucleon effects in the analysis of octant sensitivity. 
\item By incorporating the TE effect in the ESF (calculated within the RFG model) model, a notable enhancement in octant sensitivity is observed for the NH at $\delta_{CP}= 90^{\circ}$ and $\delta_{CP}= 0^{\circ}$ for both the interaction processes. Additionally, a slight improvement in sensitivity is noted for $\delta_{CP}= -90^{\circ}$ for both interaction processes.
\item In the case of NH, in terms of $\delta_{CP}$ value, among these curves, the one corresponding to $\delta_{CP}= 90^{\circ}$ exhibits higher sensitivity compared to the others, which may be due to the addition of both neutrino and antineutrino modes.
\item On the other hand, in the case of IH, all three curves overlap which implies that octant sensitivity is independent of the value of the specific value of $\delta_{CP}$ for IH.
\item (NH,$\delta_{CP}= 90^{\circ}$) with (ESF+TEM) shows octant sensitivity $> 3\sigma$ range for QE(+RPA)+2p2h while for ESF without TEM shows at < 2$\sigma$. (NH, $\delta_{CP}= -90^{\circ}$) and (NH, $\delta_{CP}= 0^{\circ}$) shows sensitivity $>2\sigma$. For all other cases, sensitivity is $<3\sigma$. 
\end{itemize}

\begin{figure*}
\centering\includegraphics[width=17cm, height=11cm]{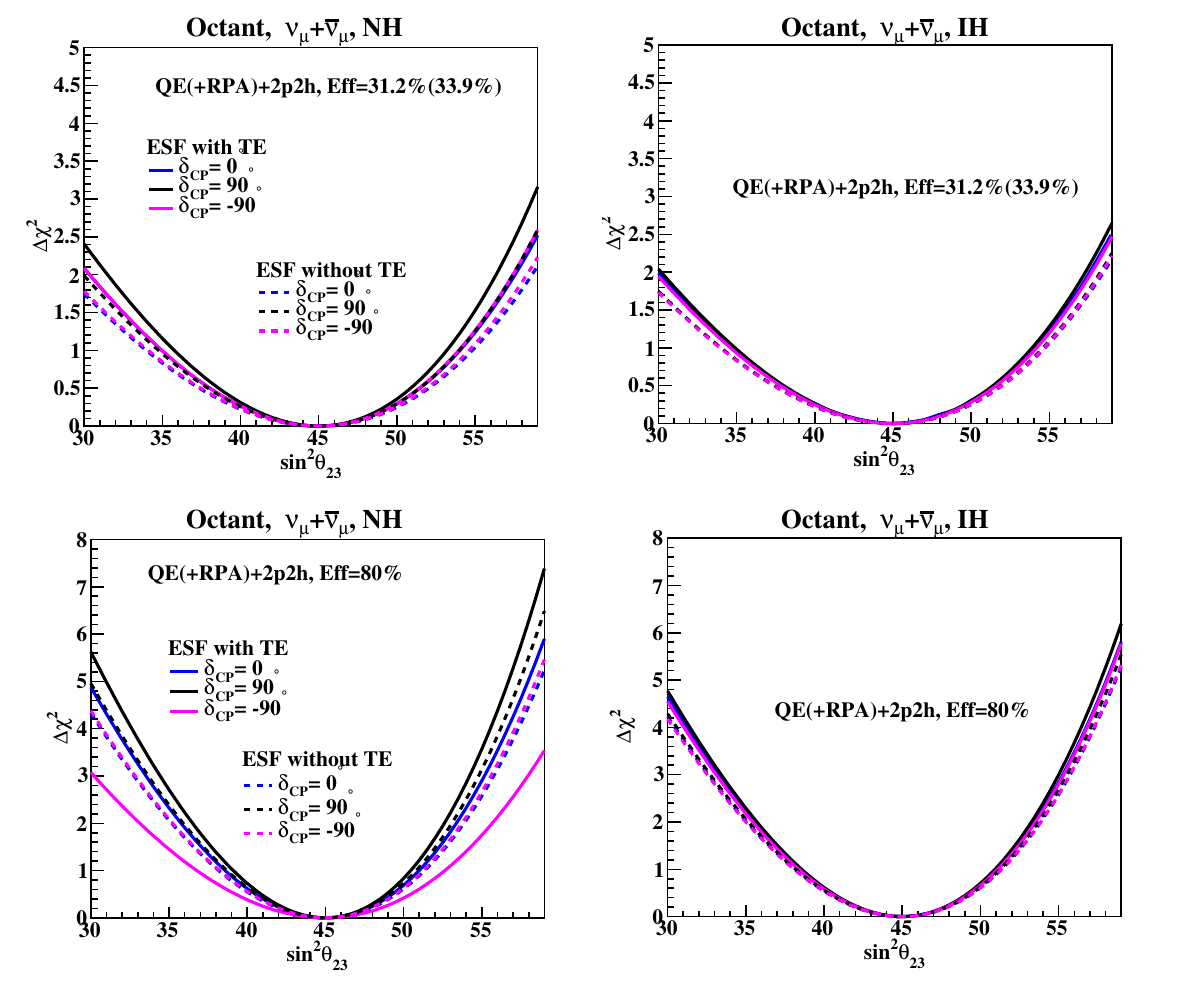}
\caption{The figure illustrates the comparison of octant sensitivity as a function of the true value of $\theta_{23}$ exclusively for the QE(+RPA)+2p2h interaction process. The ESF with TE is represented by solid lines, while the ESF without TE is indicated by dashed lines. The detector effect is considered, and the results are presented for two efficiency scenarios: 31.2\% ($\nu_{\mu}$) and 33.9\% ($\bar\nu_{\mu}$) in the top panel and 80\% efficiency in the bottom panel. The analysis is conducted for the $\nu_{\mu}$+$\bar\nu_{\mu}$ mode, with the left panel corresponding to the NH and the right panel corresponding to the IH.}
\label{chapter3}
\end{figure*}

In Fig. \ref{chapter3} we show the comparison of octant sensitivity $\chi^{2}$ as a function of the true value of $\theta_{23}$ taking into account the detector effect only for QE(+RPA)+2p2h process. The results are presented for the $\nu_{\mu}$+$\bar\nu_{\mu}$ mode for efficiency of 31.2\% ($\nu_{\mu}$) and 33.9\% ($\bar\nu_{\mu}$) and for 80\%. The left panel corresponds to NH, while the right panel corresponds to IH. Solid lines depict the ESF with TE effect, whereas dashed lines represent ESF without TE. From this figure, one can determine the range of $\theta_{23}$ values for which the octant can be determined for $\delta_{CP}=\pm90^{\circ}$ at a specific confidence level. Though detector effects enhance the discrimination among LHP and UHP of the CPV phase, however, degeneracy with respect to octant of $\theta_{23}$ is still present (better for NH as true MH), and other methods are needed to break this degeneracy \cite{Bora:2014zwa}. However, upon incorporating detector effects with an efficiency of 80\%, there is a slight increase in sensitivity in the HO for both MH, surpassing the 2$\sigma$ level. This improvement is also observed for NO$\nu$A efficiency, though remaining below the $2\sigma$ threshold. And sensitivity to measure $\theta_{23}$ is improved as detector efficiency is improved. Comparison between two panels of Fig. \ref{chapter3} shows that for IH, these sensitivities are less.

\begin{figure*}
\centering\includegraphics[width=17cm, height=11cm]{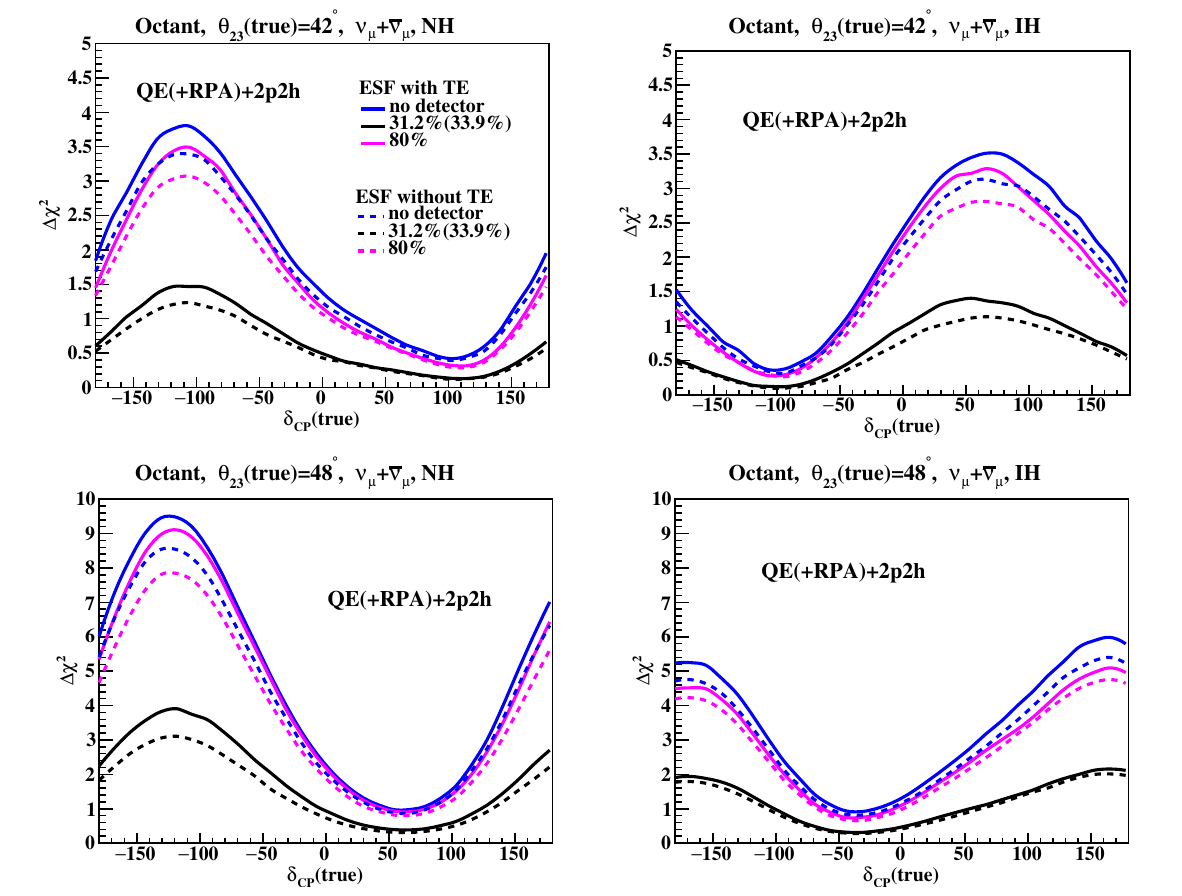}
\caption{The octant sensitivity is compared as a function of the true value of $\delta_{CP}$ for two different true values of $\theta_{23}$. The ESF with TE is represented by solid lines, while the ESF without TE is indicated by dashed lines. In the top panel, where $\theta_{23}=42^{\circ}$, and in the bottom panel, where $\theta_{23}=48^{\circ}$, the analysis is performed for both the NH in the left panel and the IH in the right panel, considering the $\nu_{\mu}$+$\bar{\nu}_{\mu}$ mode. The curves are shown for two different detector efficiencies: 31.2\% (33.9\%), represented by the black line, and 80\%, represented by the magenta line.}
\label{chapter4}
\end{figure*}

In Fig. \ref{chapter4}, we compare the octant sensitivity as a function of the true value of $\delta_{CP}$ for two different true values of $\theta_{23}$. The top panel corresponds to $\theta_{23}=42^{\circ}$ (LO), while the bottom panel corresponds to $\theta_{23}=48^{\circ}$ (HO). The analysis is performed for both the NH and IH hierarchies, displayed in the left and right panels, for the $\nu_{\mu}$+$\bar{\nu}_{\mu}$ mode. The curves are shown for two different detector efficiencies: 31.2\% (33.9\%) represented by the black line and 80\% represented by the magenta line. From this figure, it is observed that:

\begin{itemize}
\item In (LO-NH) the sensitivity improves to approximately $2\sigma$ in the LHP and reaches its peak at around $\delta_{CP}= -90^{\circ}$, and for (LO-IH) the sensitivity improves to approximately $2\sigma$ in the UHP and reaches its maximum around $\delta_{CP}= 90^{\circ}$.
\item Once again, upon introducing the TE effect into the ESF model, there is a significant improvement in sensitivity in all four scenarios. The contrast between ESF with TE and ESF without TE is particularly pronounced near the peak point.
\item (HO-NH) with no detector and 80\% efficiency shows much better sensitivity ($\geqslant3\sigma$) in the LHP for ESF+TEM while ESF without TEM, it is $\leqslant 3\sigma$. At HO-IH, sensitivity shows two peaks at $\delta_{CP}= -180^{\circ}$ and $\delta_{CP}= 180^{\circ}$. \\
The above two points can be justified as LHP is the favorable plane for NH.

\item 80\% detector efficiency shows sensitivity close to no detector effect (100\% efficiency and no resolution function) at $\geqslant2\sigma$ while NO$\nu$A efficiency shows less sensitivity $\leqslant2\sigma$ for (LO-NH) and (LO-IH).

\item (LHP-HO-NH) shows the highest sensitivity among the four panels.
\end{itemize}

\begin{figure*}
\centering\includegraphics[width=17cm, height=8cm]{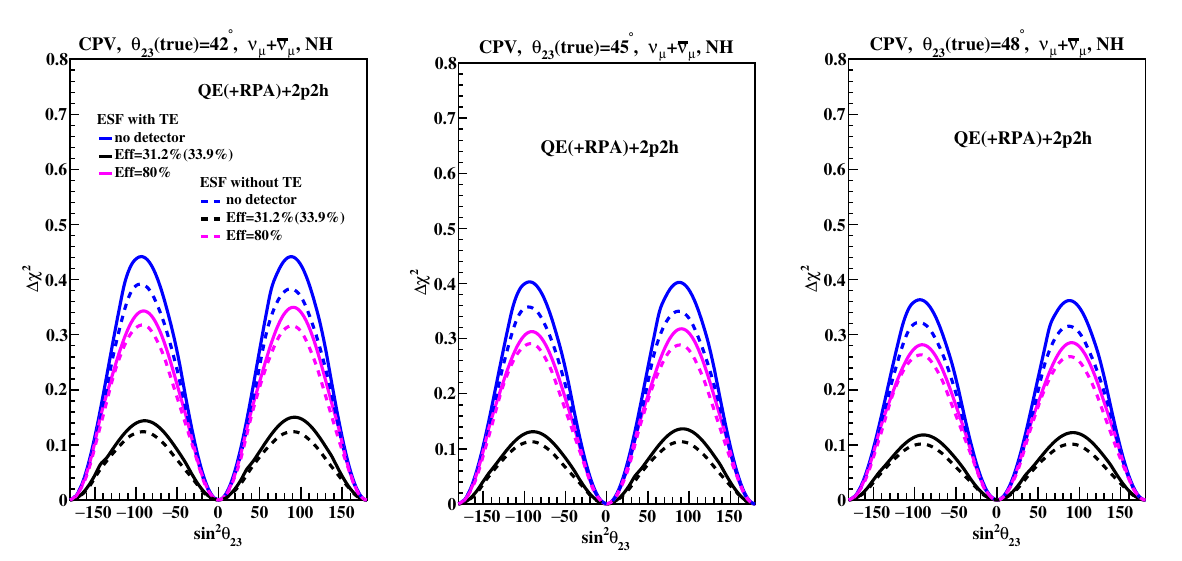}
\caption{The sensitivity comparison of CPV, as a function of $\delta_{CP}$(true), is illustrated for various true values of $\theta_{23}$ with detector effects (black and magenta lines) and without detector effects (blue line). The ESF with TE is represented by solid lines, while the ESF without TE is indicated by dashed lines. This analysis is conducted for the Normal Hierarchy in the $\nu_{\mu}+\bar\nu_{\mu}$ mode, considering the QE(+RPA)+2p2h interaction process. The left, middle, and right panel show $\theta_{23}=42^{\circ}$, $45^{\circ}$ and $48^{\circ}$.} 
\label{chapter5}
\end{figure*}

\subsection{CP violation Sensitivity}

Since the true value of $\delta_{CP}$ is unknown, the analysis involves scanning all possible true values of $\delta_{CP}$ over the range $-\pi<\delta_{CP}<+\pi$ and comparing them with CP-conserving values, such as 0 or $\pm\pi$. To detect CP violation, the value of the CP phase must differ from 0 or $\pm\pi$. To investigate CPV sensitivity, we compute and study two quantities,
 
\begin{equation}
\Delta \chi^{2}_{0}=\chi^{2}(\delta_{CP}=0)-\chi^{2}_{true}
\end{equation}
 
\begin{equation}
\Delta \chi^{2}_{\pi}=\chi^{2}(\delta_{CP}=\pi)-\chi^{2}_{true}
\end{equation}

and then we consider
 \begin{equation}
 \Delta \chi^{2}=min(\Delta \chi^{2}_{0}, \Delta \chi^{2}_{\pi})
 \end{equation}
 
 \begin{figure*}
\centering\includegraphics[width=17cm, height=8cm]{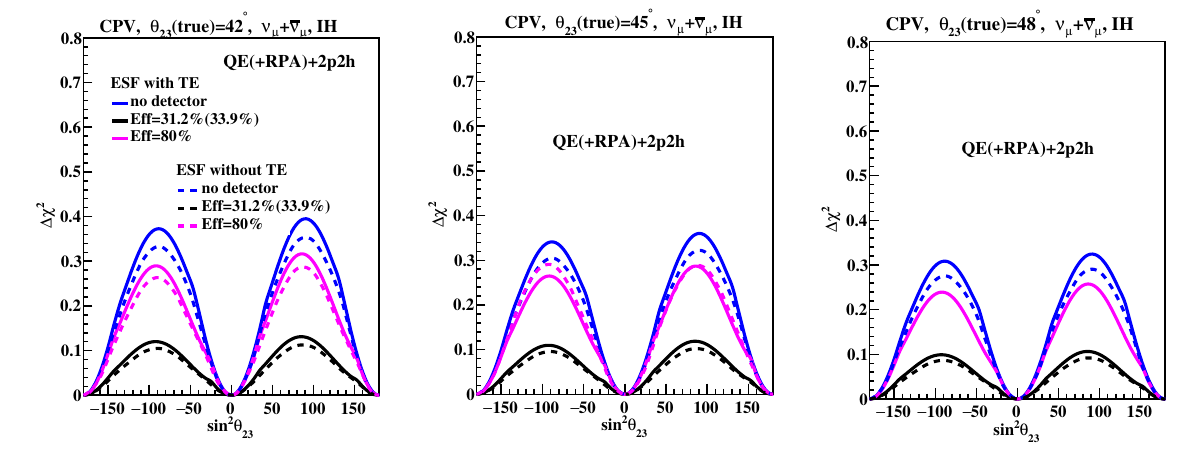}
\caption{The sensitivity comparison of CPV, as a function of $\delta_{CP}$(true), is illustrated for various true values of $\theta_{23}$ with detector effects (black and magenta lines) and without detector effects (blue line). The ESF with TE is represented by solid lines, while the ESF without TE is indicated by dashed lines. This analysis is conducted for the Inverted Hierarchy in the $\nu_{\mu}+\bar\nu_{\mu}$ mode, considering the QE(+RPA)+2p2h interaction process. The left, middle, and right panel show $\theta_{23}=42^{\circ}$, $45^{\circ}$ and $48^{\circ}$.}
\label{chapter6}
\end{figure*}

In Figs. \ref{chapter5} and \ref{chapter6}, we have presented the CPV discovery potential of the NO$\nu$A experiment. CP violation discovery potential is defined as the capability to distinguish a value of $\delta_{CP}$ from the CP conserving values of $0^{\circ}$ and $180^{\circ}$. The comparison of CPV sensitivity for different true values of $\theta_{23}$ with (black, magenta line) and without (blue line) detector effects are shown in the left, middle, and right panels for both NH (Fig. \ref{chapter5}) and IH (Fig. \ref{chapter6}). The variation of $\chi^{2}$ vs $\delta_{CP}$ (true) is presented in the left, middle, and right panel for $\theta_{23}=42^{\circ}$, $45^{\circ}$ and $48^{\circ}$. From these two figures, it is observed that

\begin{itemize}
\item For both NH and IH, $\theta_{23}=42^{\circ}$ shows slightly improved sensitivity than $\theta_{23}=45^{\circ}$ and $48^{\circ}$.
\item There is a significant difference among no detector, NO$\nu$A efficiency, and 80\% efficiency for both NH and IH.
\item The (UHP-IH) peak shows slightly more sensitivity than the (LHP-IH) peak, while in NH, the (LHP-NH) peak shows slightly more sensitivity than the (UHP-NH) peak.
\item Once more, with ESF+TE, a substantial enhancement in sensitivity is evident across all scenarios for both NH and IH. The distinction between ESF with TE and ESF without TE becomes especially noticeable near the peak point.
\end{itemize}

\subsection{Mass Hierarchy Sensitivity}

\begin{table}[h]
\begin{center}
\begin{tabular}{|c|c|c|}
\hline
parameter & best fit & Test 3 $\sigma$ range \\
\hline
$ \Delta m_{21}^2[10^{-5} eV^{2}]$ & $7.50$ & fixed\\
$ |\Delta m_{31}^2|[10^{-3} eV^{2}]$(NH) & $2.56$ & $2.46:2.65$ \\
$ |\Delta m_{31}^2|[10^{-3} eV^{2}]$(IH) & $2.46$ & $2.37:2.55$ \\
$\sin^{2}\theta_{12}/10^{-1}$ & $3.18$ & fixed\\
$ \theta_{23}/10^{-1}$(NH) & $42^{\circ}$(LO), $48^{\circ}$(HO) & $41.63^{\circ}:51.32^{\circ}$ \\
$ \theta_{23}/10^{-1}$(IH) & $42^{\circ}$(LO), $48^{\circ}$(HO) & $41.88^{\circ}:51.30^{\circ}$ \\
$ \sin^2\theta_{13}/10^{-2}$(NH) & $2.225$ & fixed \\
$ \sin^2\theta_{13}/10^{-2}$(IH) & $2.250$ & fixed \\
$\delta_{CP}$ & $-180^{\circ}:180^{\circ}$ & $0^{\circ}$(fixed)\\
\hline
\end{tabular}
\end{center}
\caption{The true and test values of the oscillation parameters used in mass hierarchy sensitivity. The best-fit values for the oscillation parameters are taken from \cite{deSalas:2020pgw}.}
\label{tab:61}
\end{table}

The hierarchy sensitivity of an experiment is determined by taking the true spectrum with the correct hierarchy and the test spectrum with the wrong hierarchy. Following formulae are used to determine the $\Delta \chi^{2}$ for mass hierarchy sensitivity:

\begin{equation}
\Delta \chi^{2}_{MH}=\chi^{2}_{IH}-\chi^{2}_{NH}
\end{equation}

\begin{equation}
\Delta \chi^{2}_{MH}=\chi^{2}_{NH}-\chi^{2}_{IH}
\end{equation}

\begin{figure*}
\centering\includegraphics[width=17cm, height=11cm]{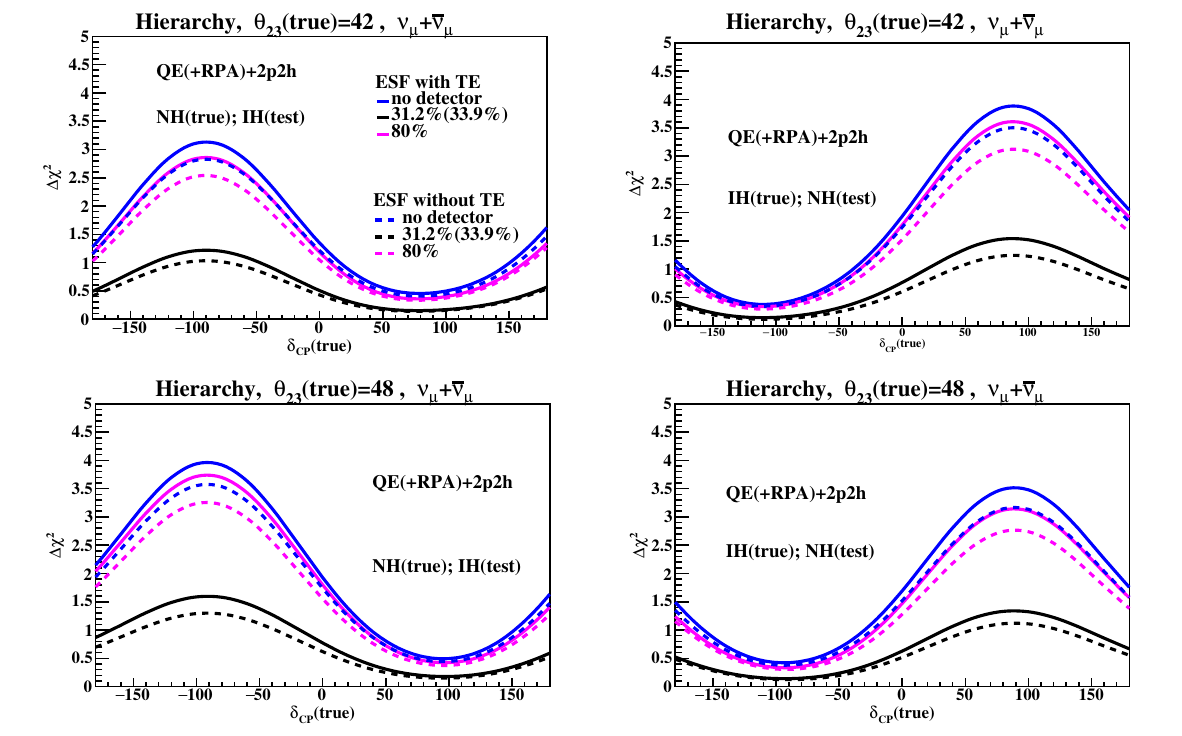}
\caption{The hierarchy sensitivity is depicted as a function of the true $\delta_{CP}$, considering both with and without detector effects for different combinations in the $\nu$+$\bar\nu$ mode. The interaction process involves QE(+RPA)+2p2h, and the ESF with TE is represented by solid lines, while the ESF without TE is indicated by dashed lines. The top panel corresponds to $\theta_{23}=42^{\circ}$, and the bottom panel corresponds to $\theta_{23}=48^{\circ}$. In the left panel, the NH is assumed as the true hierarchy, and the IH is considered as the test hierarchy. In the right panel, the IH is taken as the true hierarchy, and the NH is considered as the test hierarchy.} 
\label{chapter7}
\end{figure*}

In Fig. \ref{chapter7} and \ref{chapter8}, the hierarchy sensitivity, in terms of $\chi^2$ vs. true $\delta_{CP}$, is examined for NH (IH) as the true hierarchy and IH (NH) as the test hierarchy in the $\nu$+$\bar\nu$ mode for the QE(+RPA)+2p2h and pure QE interaction processes. The true and test values of the oscillation parameters used in mass hierarchy sensitivity are shown in Table \ref{tab:61}. From Figs. \ref{chapter7} and \ref{chapter8} it is observed that:

\begin{itemize}

\item The inclusion of multinucleon effects along with ESF+TEM has a pronounced impact on the sensitivity to the neutrino mass hierarchy, as evident from the comparison of the QE(+RPA)+2p2h (Fig. \ref{chapter7}) and QE (Fig. \ref{chapter8}) interaction processes, as compared to ESF without TEM.

\item The disparity between ESF+TEM and ESF without TEM is particularly pronounced near the peak in all four panels. Hierarchy sensitivity is observed to be within or below the $2\sigma$ range for both QE and QE with RPA and 2p2h contribution (QE(+RPA)+2p2h).

\item The hierarchy sensitivity is highest when $\delta_{CP}=-90^{\circ}$ and lowest when $\delta_{CP}=90^{\circ}$ when NH is the true hierarchy and IH is the test hierarchy. Conversely, the opposite behavior is observed in the left panel when IH is the true hierarchy and NH is the test hierarchy.

\item For true MH NH, $\delta_{CP}= -90^{\circ}$ and true MH IH, $\delta_{CP}= 90^{\circ}$ sensitivity $\leqslant2\sigma$ but sensitivity is $<1\sigma$ for true MH NH, $\delta_{CP}= 90^{\circ}$ and true MH IH, $\delta_{CP}= -90^{\circ}$ for both QE(+RPA)+2p2h and pure QE. It is evident that in the favorable half-plane (LHP-NH and UHP-IH) of $\delta_{CP}$ values, the ability to differentiate the correct neutrino mass hierarchy is significantly enhanced. This indicates that the sensitivity to determine the true hierarchy is greater when $\delta_{CP}$ falls within the favorable half-plane. 

\item Similar observations are noted in the case of QE interactions (Fig. \ref{chapter8}) with both mass hierarchy sensitivities falling within the $2\sigma$ range. It is evident that the inclusion of multinucleon+ RPA with ESF+TEM leads to a substantial increase in sensitivity.

\begin{figure*}
\centering\includegraphics[width=17cm, height=11cm]{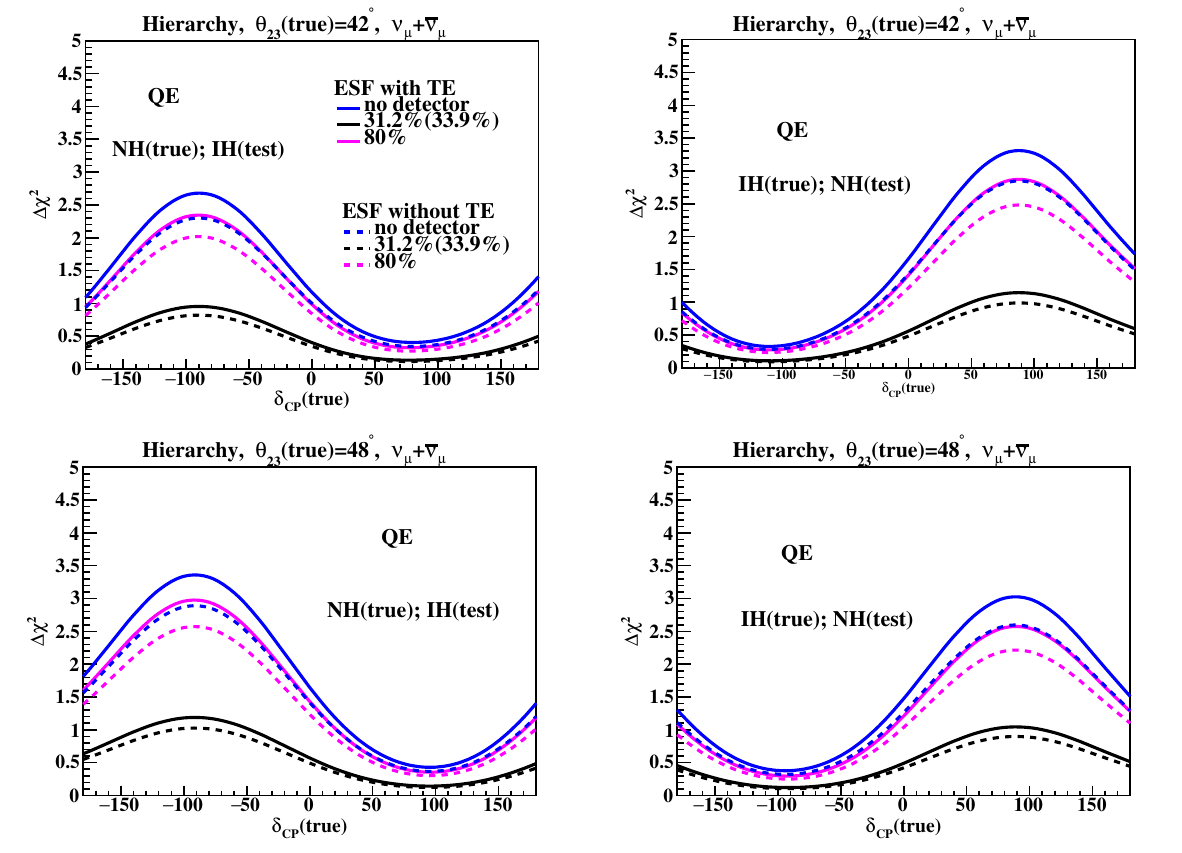}
\caption{The hierarchy sensitivity is depicted as a function of the true $\delta_{CP}$, considering both with and without detector effects for different combinations in the $\nu$+$\bar\nu$ mode. The interaction process involves QE, and the ESF with TE is represented by solid lines, while the ESF without TE is indicated by dashed lines. The top panel corresponds to $\theta_{23}=42^{\circ}$, and the bottom panel corresponds to $\theta_{23}=48^{\circ}$. In the left panel, the NH is assumed as the true hierarchy, and the IH is considered as the test hierarchy. In the right panel, the IH is taken as the true hierarchy, and the NH is considered as the test hierarchy.} 
\label{chapter8}
\end{figure*}

\item For (true MH NH, HO, LHP) and (true MH IH, LO, UHP) cases, the hierarchy sensitivity can reach close to $2\sigma$ for the complete model.

\item The sensitivity to determine the MH shows a slight advantage in the higher octant for true NH and lower octant for true IH. 
\end{itemize}

\renewcommand{\arraystretch}{4}
\begin{table}[h]
\begin{center}
\scalebox{0.88}{
\begin{tabular}{|c|c|c|}
\hline
\textbf{Octant Sensitivity (ESF+TE)} & \textbf{CPV Sensitivity (ESF+TE)} & \textbf{MH Sensitivity (ESF+TE)} \\
\hline
\makecell{With no detector effects for \\NH $\delta_{CP}= 90^{\circ}$ shows\\ maximum octant sensitivity, while \\for IH there is no distinction \\among curves for \\$\delta_{CP}= 0^{\circ}, 90^{\circ}$, and $-90^{\circ}$} & \makecell{LO shows slightly\\ better CPV sensitivity for \\both MH than HO}		&	\makecell{For MH NH, LHP shows \\better sensitivity than\\ UHP, i.e., sensitivity to \\determine true MH is better \\when $\delta_{CP}$ falls within\\ favorable half plane}\\
\hline
\makecell{Better detector \\efficiency implies better \\octant sensitivity} & \makecell{UHP-IH shows slightly \\better CPV sensitivity than\\ LHP-IH, while in NH, the\\ (LHP-NH) peak shows \\ slightly more sensitivity \\ than the (UHP-NH) peak} & \makecell{For true NH, LHP-HO \\and true IH, UHP-LO case \\MH sensitivity can reach \\close to $\sim 2\sigma$ significance}\\
\hline
\makecell{Octant sensitivity for \\UHP-HO-NH is $\geqslant3\sigma$ \\and for LHP-LO-NH $\geqslant2\sigma$} & \makecell{Significant difference among \\no detector, NO$\nu$A efficiency \\and 80\% efficiency for \\both NH and IH}	&	\makecell{The sensitivity is slightly \\higher in HO-NH, and \\LO-IH cases}\\
\hline
\end{tabular}}
\end{center}
\caption{Summary of the work done in this work.}
\label{tab:62}
\end{table}

\section{Summary and Conclusion}
\label{sec:4}

In this work we investigated the potential for determining the octant of the atmospheric mixing angle $\theta_{23}$, the CP-violating phase $\delta_{CP}$, and the sensitivity to the mass hierarchy in the appearance channel of NO$\nu$A when multinucleon and detector effects are included. The concept of octant degeneracy traditionally refers to the ambiguity between $\theta_{23}$ and $\pi/2-\theta_{23}$. We conducted a $\chi^{2}$ analysis focusing on the octant of $\theta_{23}$, $\delta_{CP}$, and hierarchy sensitivity. Moreover, we have studied the comparison between the nuclear model's Effective Spectral Function (calculated within the RFG model) with transverse enhancement and ESF without transverse enhancement in terms of sensitivity analysis. The findings demonstrated that incorporating multinucleon and detector effects, alongside ESF with TEM (ESF+TEM), significantly enhances sensitivities for octant determination, CP phase measurement, and Mass Hierarchy identification, in contrast to ESF without TEM. We also find that between the two interaction processes, QE(+RPA)+2p2h shows significantly improves sensitivity as compared to only QE case, and we highlighted their impact on the extraction of neutrino oscillation parameters in LBL experiments.\\

 After a careful analysis, the results of this work can be summarised as shown in Table \ref{tab:62}.

\renewcommand{\arraystretch}{4}
\begin{table}[h]
\begin{center}
\scalebox{0.88}{
\begin{tabular}{|c|c|c|}
\hline
Preference of Octant Sensitivity & Preference of CPV Sensitivity & Favored MH Sensitivity \\
\hline
\makecell{LBL gives two \\degenerate solutions for\\ both LO and HO} & \makecell{Values of CPV phase\\in the vicinity of $\delta_{CP}=\pi/2$\\are excluded by 3$\sigma$ \\for IH, values around $\delta_{CP}=3\pi/2$\\ in NH are disfavored at 2$\sigma$}		&	\makecell{Independent analysis of \\both T2K and NO$\nu$A does \\not show specific preference \\of MH}\\
\hline
\makecell{LBL+ATM and LBL+\\Reactor shifts towards \\HO} & \makecell{From LBL+reactor data\\CP conserving value\\$\delta_{CP}=0$ is disfavored\\but $\delta_{CP}=\pi$ is still allowed} & \makecell{As a consequence of \\tension in T2K and NO$\nu$A \\data all LBL data favor IO \\over NO}\\
\hline
\end{tabular}}
\end{center}
\caption{Summary of the latest values for oscillation parameters and the favored octant of $\theta_{23}$, $\delta_{CP}$, and the mass hierarchy obtained from multiple global experiments \cite{NOvA:2021nfi, deSalas:2020pgw}.}
\label{tab:63}
\end{table}

Our findings revealed that the higher octant of $\theta_{23}$, the lower half plane of $\delta_{CP}$, and the normal mass hierarchy (HO-LHP-NH) \textcolor{red}{with ESF+TEM} exhibit improved sensitivity, enabling a more precise determination of the corresponding parameters. The analysis using our comprehensive model QE(+RPA)+2p2h with ESF+TEM demonstrated significantly enhanced sensitivity compared to the pure QE interaction process. Furthermore, it is worth noting that the curves representing 80\% efficiency (overestimated) closely align with the curves corresponding to no detector effects. These no-detector effect curves represent an ideal scenario with a 100\% efficient detector and no consideration of a resolution function. On the other hand, the curves obtained with 31.2\% efficiency for $\nu_{\mu}$ and 33.9\% efficiency for $\bar\nu_{\mu}$ display more noticeable deviations from the no detector effect curves.\\

In conclusion, incorporating multinucleon effects and improving detector efficiency, which constitutes the novelty of the research presented in this paper, have the potential to enhance the capabilities of the NO$\nu$A (and other long-baseline) experiments in conducting precise parameter studies. This approach also provides deeper insights into the significant issue of parameter degeneracies present in the measurements of long-baseline neutrino experiments. Hence, the appropriate nuclear models, multinucleon (QE(+RPA)+2p2h with ESF+TEM), and detector effects should be carefully included in the analyses of LBL experimental data, for precise determination of unknown neutrino oscillation parameters and for resolving parameter degeneracies as well. The results presented in this work can be used in improving the nuclear models in future related analyses, and they can help improve our understanding of some unsolved crucial physics issues such as baryon asymmetry of the universe (BAU), the contribution of neutrinos to dark matter, etc.

\end{document}